\documentclass[onecolumn,review,12pt]{elsarticle} 

\usepackage{amsmath,amssymb,amsfonts,amsthm,makeidx,graphicx}
\usepackage{txfonts}
\usepackage{helvet}

\journal{Journal of Nuclear Materials}

\begin{document}

\begin{frontmatter}

\title{Amorphous High-Density Plutonium}

\author[1]{Jonathan I. Katz\corref{cor1}}%
\ead{katz@wuphys.wustl.edu}

\author[2]{Anthony Rollett}%
\ead{rollett@andrew.cmu.edu}

\author[3]{Russell J. Hemley}%
\ead{rhemley@uic.edu}

\cortext[cor1]{Corresponding author}

\affiliation[1]{organization={Department of Physics and McDonnell Center for the Space Sciences, Washington University},
addressline={\\ 1 Brookings Dr.},
postcode={63130},
city={St. Louis, Mo.},
country={U. S. A.}}

\affiliation[2]{organization={Deptartment of Materials Science and Engineering, Carnegie-Mellon University},
addressline={5000 Forbes Ave.},
postcode={15213},
city={Pittsburgh, Penn.},
country={U. S. A.}}

\affiliation[3]{organization={Departments of Physics, Chemistry and Earth and Environmental Sciences, University of Illinois},
addressline={845 W. Taylor St.},
postcode={60607},
city={Chicago, Ill.},
country={U. S. A.}}



\begin{abstract}[Abstract]
	Metastable aluminum-alloyed $\delta$-plutonium shrinks rapidly and
	pure $\alpha$-plutonium swells rapidly at 4 K.  At ambient
	temperature alloyed $\delta$-plutonium slowly swells, but its bulk
	density decreases more slowly than would be inferred from the
	increase in its lattice parameter.  These results might be
	explained as the result of ingrowth of the opposite phases, but they
	have not been found in X-ray diffraction.  The cryogenic results may
	be explained by ingrowth of an amorphous phase with density
	intermediate between those of $\alpha$ and $\delta$.  This phase
	rapidly but not instantaneously anneals to its parent phases at
	temperatures $\gtrapprox 100\,$K.
\end{abstract}

\begin{highlights}
\item Rapid shrinkage of $\delta$-plutonium and swelling of $\alpha$-plutonium at 4 K indicates formation of an intermediate density amorphous phase.
\item Discrepancy between growth of lattice parameter and of length of $\delta$-plutonium at room temperature indicates formation and delayed annealing of a denser phase.
\item Absence of $\alpha$-plutonium in X-ray diffraction of aging $\delta$-plutonium indicates the denser phase is amorphous.
\end{highlights}


\end{frontmatter}

\newpageafter{abstract}

\section{Introduction}\label{introduction}
The equilibrium phase of plutonium at room temperature is the dense ($\rho
= 19.86$ g/cm$^3$) monoclinic $\alpha$ phase.  The thermodynamically
unstable low density ($\rho = 15.92$ g/cm$^3$) face centered cubic $\delta$
phase has been the subject of most study; it is stabilized kinetically (and
partially thermodynamically) by alloying with a few atomic percent of
aluminum or gallium.

All isotopes of plutonium are radioactive, the most important ones by
$\alpha$-decay.  Widely studied isotopic mixtures, typically 94\%
$^{239}$Pu, introduce helium at a rate of about 40 ppm (atomic) per year.
The ingrowth of helium, and of $^{241}$Am from $\beta$ decay of traces of
$^{241}$Pu, are predicted to produce very slow swelling, initially at a
volumetric rate of about $5 \times 10^{-5}$/y \cite{W06}.
The $\alpha$ particle and recoiling uranium nuclei resulting from these
decays damage the plutonium lattice at a rate of about 0.1 displacement per
atom (dpa) per year in $^{239}$Pu, and at higher rates in material doped
with highly radioactive ($t_{1/2} = 87.8\,$y) $^{238}$Pu.  This damage
disorders the lattice and changes its properties in ways that are not fully
understood and that are sensitive to temperature.

The low-temperature behavior of plutonium undergoing radioactive decay and
self-irradiation has long been known to be unusual.  At temperatures
$> 15\,$K \cite{O16}, but not at $4\,$K \cite{MM71}, the damaged material
anneals at temperature-dependent rates.  Annealing is also inferred from the
presence of transients in $\delta$-Pu at ambient temperature \cite{C06,R07},
the reversibility of its rapid shrinkage at 4 K upon annealing at 97 K and
the reversibility of the rapid swelling of $\alpha$-Pu at 4 K upon annealing
at 92 K \cite{M70}.

In this letter {we follow \cite{MH72} in suggesting that radiation
damage produces} a distinct, metastable, non-equilibrium phase, {with
density intermediate between that of the $\alpha$ and $\delta$ phases,}
disordered enough that it does not show X-ray diffraction peaks.  At ambient
temperature in $\delta$-Pu it relaxes to the crystalline $\delta$-phase
almost as fast as it is formed, manifesting itself as a small discrepancy
between the growth of lattice parameter and the swelling of a macroscopic
sample.  At 4 K, where kinetic barriers preclude relaxation, over times of
thousands of hours it accounts for several percent volumetric shrinkage of
$\delta$-Pu and several percent volumetric swelling of $\alpha$-Pu.
\section{Cryogenic Behavior}
Most measurements of density changes as plutonium self-irradiates have been
made at room temperature, but studies at 4 K \cite{M70,MH72,JL,JLalpha} were
reviewed by \cite{H08}\footnote{{\it N.B.\/} The densities in Tables I of
\cite{MH72,H08} are extrapolations to infinite time.}.  At such low
temperature relaxation and annealing are expected to be negligible (a report
\cite{JLalpha} that the behavior of $\alpha$-Pu 1.5 K differs from that at
4 K was later disputed \cite{M70}).

At 4 K $\delta$-Pu initially {\it shrinks} rapidly; in 0.5 year a
sample shrank by about 1.4\% in length at a gradually decreasing rate that
would extrapolate to an asymptotic volumetric shrinkage of about 15\% after
a few years \cite{M70,JL}.  This is nearly $2000$ times as fast as it swells
at ambient temperature \cite{C06}.  Most of these data describe length
shrinkage; there are a few crystallographic data \cite{M70} at very early
time ($< 0.01$ dpa) that appear approximately consistent with the rate of
length contraction.  

Less attention has been paid to $\alpha$-Pu, but at 4K \cite{M70,MH72,H08}
reported an initial linear growth rate of about 3\% per year, decaying with
an $e$-folding time of about a year.  This is comparable in magnitude, but
opposite in sign, to the shrinkage rate found for $\delta$-Pu \cite{M70}.
It extrapolates to almost the same asymptotic density as does
$\delta$-plutonium \cite{M70,MH72,H08}.  Both shrinkage and expansion were
almost entirely reversed upon annealing at about 100 K \cite{M70,JLalpha},
consistent with the formation of a metastable state of density intermediate
between those of the $\delta$ and $\alpha$ phases.


The rates of density change of both phases of Pu gradually decrease,
suggesting convergence to a phase of intermediate density in a time of about
a year.  This is explained if the material, whether initially $\delta$ or
$\alpha$, is largely converted to the intermediate phase, disordered enough
not to show X-ray diffraction peaks, after that time, and that further
radiation damage does not change it.

After the roughly half year of irradiation at 4K in the experiments reported
in \cite{M70,JL} much of the $\delta$ or $\alpha$ remained, and annealing
returned the material to its original phase.  The remaining $\delta$ or
$\alpha$ may have nucleated regrowth of the original phase.  The experiments
did not extend long enough for conversion of all the material to the
intermediate phase.  It is unclear to which phase annealing would return
after complete conversion, if that occurs.

Measurements of the swelling of $\alpha$-U doped with $^{238}$Pu to achieve
a level of $\alpha$ activity about 3/2 of that of $^{239}$Pu were also
reported \cite{M70}, yet it expanded only about 40\% as fast as did
$^{239}$Pu.
This implies that less than half the expansion of $\alpha$-Pu may be
attributed directly to radiation damage (which occurs for both elements);
the remainder is specific to Pu, which we attribute to conversion of the
dense $\alpha$-Pu phase to a phase of intermediate density.

These results might be explained if radiation damage at 4 K converted
$\delta$-Pu to $\alpha$-Pu and $\alpha$-Pu to $\delta$-Pu, with saturation
at a few tenths of a dpa.  Complete conversion of $\delta$ to $\alpha$ would
increase the density by 25\%, consistent with the extrapolation of the
early-time cryogenic density increase to longer times.  However, conversion
of $\alpha$-Pu to $\delta$-Pu by radiation damage is implausible; 
$\alpha$-Pu is stable at low temperatures, while $\delta$-Pu is only
kinetically stabilized.  Conversion of $\delta$-Pu to $\alpha$-Pu is
thermodynamically possible, if radiation damage overcomes the kinetic
barrier, but this is inconsistent with reversal of the contraction upon
annealing---$\alpha$-Pu cannot anneal to $\delta$-Pu below the
$\gamma$-$\delta$ transition temperature around 600 K.

Radiation damage might prefer one phase to the other, but it is implausible
that it would always prefer the ``other'' phase; having converted one to the
other, this would require that radiation damage would then begin to reverse
the process.  There have been no reports of $\alpha$ structure in the X-ray
diffraction of contracted $\delta$-Pu, or {\it vice versa\/}, so we reject
this hypothesis.

These experimental results should be compared to theoretical predictions of
swelling as a result of the ingrowth of helium and (at ages of $\lessapprox
20$ years) of decay ($t_{1/2} = 14.4$ y) of $^{241}$Pu to larger $^{241}$Am
atoms \cite{W06}.  This theory predicts a volume growth rate (at zero age)
of about 0.005\% per year, about three orders of magnitude less, and of the
opposite sign, than what was observed for $\delta$-plutonium at 4 K
\cite{M70} where no annealing is expected.

{\it Ab initio\/} calculations \cite{SSGS} of void swelling predict that it
is small in $\delta$-Pu but large in $\alpha$-Pu.  This is consistent with
ambient temperature measurements of $\delta$-Pu, but no ambient temperature
data are available for $\alpha$-Pu.  It does not explain the shrinkage of
$\delta$-Pu at 4 K (which is why we hypothesize a denser disordered phase); 
it may be qualitatively consistent with the rapid swelling of $\alpha$-Pu at
4K (\cite{SSGS} do not quantify this) but may not predict the reversal of
this swelling around 100 K.
\section{Amorphous Phase?}
We {follow \cite{MH72}} in suggesting that another phase, sufficiently
disordered not to show X-ray diffraction peaks, with density intermediate
between those of $\alpha$ and $\delta$, is formed by radiation damage under
cryogenic conditions.  It would not show the crystallographic signature of
either the $\alpha$ or the $\delta$ phase, but partial transformation to it
would shrink $\delta$ and expand $\alpha$, as observed.

Molecular dynamic calculations of radiation damage cascades in unalloyed
$\delta$-Pu indicate the formation of such a glassy phase as radiation
damage relaxes \cite{K07}.  At 300 K this amorphous phase persisted for
the entire 2 ns duration of the calculation; it is impossible to say if it
would persist on laboratory time scales, either at 300 K or at 4 K.  These
authors did not report its density, but if it were intermediate between
those of the $\alpha$ and $\delta$ phases it could explain the observed
contraction of $\delta$-Pu at 4 K and the discrepancy between lattice
parameter and dilatometric densities discussed in Sec.~\ref{roomT}.  At 180
K these calculations indicated complete transformation of the computational
domain to the glassy phase.  This has not occurred in any experiment,
perhaps because experiments on the $\delta$ phase at room or cryogenic
temperatures necessarily involve alloy-stabilized plutonium rather than the
pure element studied in \cite{K07}.

Such an amorphous phase would be metastable, and expected to anneal to the
equilibrium $\alpha$ phase at temperatures comparable to those at which
later EXAFS studies reported annealing of radiation damage in $\delta$-Pu at
temperatures 15--150 K \cite{O16}.  Annealing at temperatures in the range
90--100 K was reported \cite{M70} to reverse the density changes both
$\alpha$- and $\delta$-Pu underwent at 4 K, {supporting this hypothesis}.

Annealing of expanded $\alpha$-Pu to the density of the undamaged $\alpha$
phase is easy to understand---$\alpha$ is the equilibrium phase.  The
annealing of contracted $\delta$-Pu back to its undamaged density is more
difficult to understand.  It cannot be explained if contraction is the
result of conversion to the equilibrium $\alpha$ phase, because $\alpha$
cannot spontaneously convert to any other phase at temperatures below $\sim
400$ K.  The annealing of $\delta$-Pu irradiated at 4 K back to $\delta$ can
only be explained if radiation damage to $\delta$-Pu at 4 K produces another
metastable phase that is even more thermodynamically disfavored than is
$\delta$-Pu.

That leaves the question of why the hypothesized amorphous phase formed from
radiation damaged $\delta$ anneals to $\delta$, rather than to the
thermodynamically preferred $\alpha$.  A possible explanation is that
recrystallization nucleates on the remaining $\delta$ in contact with the
amorphous phase, and as a result assumes its crystallographic symmetry.
Conversion of $\delta$ to $\alpha$ is kinetically inhibited at these
temperatures, although it does occur at somewhat higher temperatures
\cite{K03}.  This question could be addressed by long-term X-ray diffraction
studies of cryogenic $\delta$-Pu (and particularly accelerated-aged
$\delta$-Pu, spiked with $^{238}$Pu to increase the rate of radiation
damage \cite{H3}), when nearly complete conversion to the amorphous phase
might be expected.
\section{Transients in Room Temperature Expansion}
\label{roomT}
At room temperature $\delta$-Pu very slowly swells, following
somewhat faster initial transients \cite{C06,R07}.   The fact that the rate
of swelling decreases implies that some of the radiation damage is reversed
by annealing with a relaxation time ${\cal O}(1\,$y) (0.1 dpa).  An
irreversible part produces a long term trend approximately consistent with
theoretical predictions for the ingrowth of helium and americium \cite{W06}.

X-ray diffraction indicates the $\delta$-Pu lattice expands about three
times more than the change in dimensions measured by dilatometry
\cite{C06,R07,B07,H2}.  This discordance can be explained if some of the
low density crystalline $\delta$ phase transforms to a higher density phase.
Were this higher density phase crystalline $\alpha$-Pu, it would show its
distinctive signature in X-ray diffraction, but that has not been reported.
This is explained if the higher density material is sufficiently disordered.
Alternatively, \cite{W06a} suggested that the transient increase of lattice
parameter at ambient temperature results from radiation-induced diffusion of
Ga to Pu$_3$Ga ($\zeta^\prime$ phase), with increasing lattice parameter of
the $\delta$-Pu phase as it is gradually depleted of Ga; this saturates at
about 0.1 dpa because of radiation-induced disordering.  The lesser growth
of dimension is then explained by ingrowth of Pu$_3$Ga.

The transient relaxes to slower growth at about $10^{-3}$ lattice expansion
in normally aging material \cite{R07}, but at about $2 \times 10^{-4}$
linear dilation in accelerated-aged material \cite{C06}, in both cases at
about 0.1 dpa (unfortunately, these experiments differed in two
variables---the quantities measured and the rate of radiation damage,
clouding direct comparisons).  Again, the greater expansion of lattice
parameter is partly compensated by collapse of some portion of the
(expanded) $\delta$ phase to denser material, that we suggest was amorphous.
After the initial relaxation time, further production of denser material is
balanced by annealing of existing denser material to $\delta$-Pu, and
the subsequent slow swelling results from helium and americium ingrowth.
The fact that the transients relax at similar damage, measured in dpa,
rather than at similar physical ages (the accelerated-aging material
accumulates damage 15--20 times faster, depending on the unstated $^{241}$Am
content) indicates that either the relaxation processes or the nature of
damage depend on the damage rate, calling into question the equivalence of
accelerated and natural aging.
\section{Relation to Other Data}
\begin{figure}
\centering
\includegraphics[width=0.49\columnwidth]{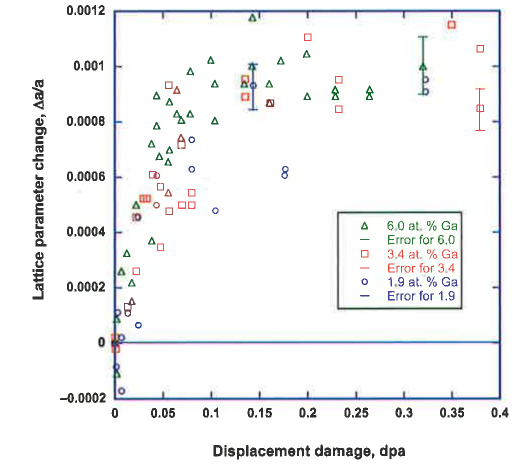}
\includegraphics[width=0.49\columnwidth]{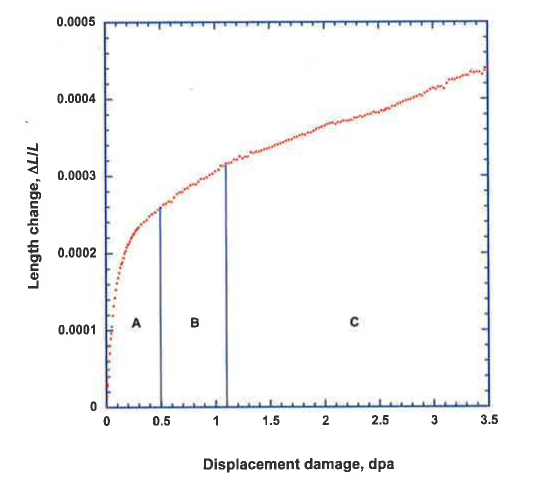}
\caption{\label{comparison}Left (from \cite{H21}): The increase of the
lattice constant of Ga-stabilized $\delta$-Pu under self irradiation.  Right
(from \cite{H22}): Dilatometry shows the much slower growth of a sample
dimension.  Note different ordinate scales and slowing of growth of both
parameters after initial transients, indicating ingrowth of denser material
and modification $\delta$-phase structure.}
\end{figure}

It is well known that X-ray diffraction of self-irradiated Ga-stabilized
$\delta$-Pu shows much more rapid growth of lattice parameter than of bulk
dimensions.  This is illustrated in Fig.~\ref{comparison}.
The sample dimension initially grows an order of magnitude more slowly than
the lattice constant, implying that some of the low-density $\delta$-Pu is
converted to a higher density phase.  The absence of diffraction evidence
for the $\alpha$ phase suggests the formation of a disordered phase that
shows no diffraction peaks, as indicated by the cryogenic data discussed in
this paper.

This higher density (compared to $\delta$-Pu) phase without diffraction
peaks may perhaps be identified with the $\sigma$ phase suggested in
\cite{C14a,C14b}.  These authors found the $\sigma$ phase to occur in
regions depleted in Ga, consistent with our suggestion that it also occurs
in pure $\alpha$-Pu at 4 K.  At room temperature this phase may rapidly
relax to $\delta$-Pu, so that the volume does not shrink as rapidly as at
cryogenic temperature because the fraction of the higher density phase
saturates after an initial transient.
\section{Discussion}
The recovery of the $\delta$-Pu aged at 4 K to its unaged lower density
upon annealing implies that the denser component induced by radiation damage
cannot be the $\alpha$ phase because thermodynamics precludes spontantaneous
transformation of $\alpha$ to $\delta$ under these conditions.  Recovery can
only be explained if the higher density phase is a nonequilibrium phase,
even further from equilibrium than $\delta$-Pu.  Such a phase may store
energy sufficient to drive it back to the $\delta$ phase once warmed enough
to overcome kinetic barriers.  We suggest that contact with surrounding
$\delta$-Pu nucleates the transformation back to $\delta$-Pu, despite the
stronger thermodynamic drive to the equilibrium $\alpha$ phase.

Many of the experimental results discussed here are quite old ({\it e.g.\/}
\cite{M70,JL,JLalpha}), and it would be desirable to repeat these cryogenic
experiments, extending them to times of several years but with particular
attention paid to the first months and year.  X-ray diffraction data should
be obtained along with dilatometry to test the inference that dimensional
changes are not the result of transformation of $\delta$-Pu to $\alpha$-Pu,
and {\it vice versa\/}, but rather imply the existence of a disordered phase.

The behavior of normally and accelerated aging material should also be
directly compared, in order to test the assumption that accelerating aging
with $^{238}$Pu faithfully accelerates the clock without changing aging in
any other way.  The rapid shrinkage of $\delta$-Pu and swelling of
$\alpha$-Pu at 4 K enable accurate measurement and comparison of their rates
in both unaccelerated and accelerated aging material, and quantitative tests
of this assumption.  At the high rate of radiation damage of pure $^{238}$Pu
several different crystal phases co-exist \cite{GS}, implying that such
extreme rates of radiation damage do more than accelerate the aging clock.

The formation of a disordered phase in plutonium as a result of
autoirradiation may be analogous to the formation of amorphous phases of
silica as a result of applied pressure \cite{K93,H94}.  In each case an
external source of free energy drives an equilibrium crystalline phase to a
non-equlibrium disordered phase.

{The interpretation of the density changes of $\alpha$- and
$\delta$-plutonium at 4 K as the formation of an amorphous phases may be
tested empirically.  Metallic glasses show \cite{CuZr} low temperature
specific heat excesses analogous to those of network and molecular glasses,
but measuring this in plutonium may be difficult.   Amorphous metals have
electrical resistivities, straightforward to measure, greater than those of
crystalline phases \cite{CuZrTi}.  Resonant ultrasound spectroscopy can
provide accurate measurements of elastic constants \cite{Migliori} to test
the hypothesis of ingrowth of an amorphous phase with self-irradiation.}
\section*{Acknowledgements}
ADR and RJH acknowledge partial support of the DOE/NNSA through the
Chicago/DOE Alliance Center (DE-NA0004153) for this work.  We thank Don
Brown for useful discussions and pointing out \cite{MH72} and Saryu Fensin
for discussions of experimental feasibility.
\bibliographystyle{elsarticle-num}
\bibliography{Puamorph} 

@ARTICLE{JL,
	title = "Contraction des phases $\beta$ et $\delta$ du plutonium sous l'effet de l'autoirradiation a 4.2$^\circ$ {K}?",
	author = "Jacquemin, J. and Lallement, R.",
	journal = "Physics Letters A",
	volume = "33",
	number = "6",
	pages = "384--385",
	year = "1970"
}

@ARTICLE{JLalpha,
	title = "Allongement par autoirradiation a 1.5$^\circ$ K du plutonium $\alpha$",
	author = "Jacquemin, J. and Lallement, R.",
	journal = "Physics Letters A",
	volume = "32",
	number = "3",
	pages = "181--182",
	year = "1970"
}

@ARTICLE{C06,
	title = "Density changes in plutonium observed from accelerated aging using {P}u-238 enrichment",
	author = "Chung, B. W. and Thompson, S. R. and Woods, C. H. and Hopkins, D. J. and Gourdin, W. H. and Ebbinghaus, B. B.",
	journal = "J. Nucl. Matl.",
	volume = "355",
	pages = "142--149",
	year = "2006"
}

@ARTICLE{R07,
	title = "Study by {XRD} of the lattice swelling of {P}u{G}a alloys induced by self-irradiation",
	author = "Ravat, B. and Oudot, B. and Baclet, N.",
	journal = "J. Nucl. Mater.",
	volume = "366",
	pages = "288--296",
	year = "2007"
}

@CONFERENCE{M70,
	title = "Length Changes in Actinide Metals Due to Self-Irradiation Damage at 4 {K}",
	author = "Marples, J. A. C. and Hough, A. and Mortimer, M. J. and Smith, A. and Lee, J. A.",
	booktitle = "Plutonium 1970 and Other Actinides",
	publisher = "The Metallurgical Society of the AIME",
	address = "Englewood, Col.",
	pages = "635--648",
	editor = "Miner, W. N.",
	series = "Proc. Fourth Int. Conf. on Plutonium and Other Actinides",
	year = "1970"
}

@ARTICLE{K07,
	title = "Modeling of aging in plutonium by molecular dynamics",
	author = "Kubota, A. and Wolfer, W. and Valone, S. and Baskes, M. J.",
	journal = "J. Comput.-Aided Mater. Des.",
	volume = "14",
	pages = "367--378",
	year = "2007"
}

@ARTICLE{W06,
	title = "Volume changes in $\delta$-plutonium from helium and other decay products",
	author = "Wolfer, W. G. and S{\"o}derlind, P. and Landa, A.",
	journal = "J. Nucl. Matl.",
	volume = "355",
	pages = "21--29",
	year = "2006"
}

@ARTICLE{W06a,
	title = "Reversible expansion of gallium-stabilized $\delta$-plutonium",
	author = "Wolfer, W. G. and Oudot, B. and Baclet, N.",
	journal = "J. Nucl. Matl.",
	volume = "359",
	pages = "185--191",
	year = "2006"
}

@ARTICLE{O16,
	title = "Isochronal annealing effects on local structure, crystalline fraction and undamaged region size of radiation damage in {G}a-stabilized $\delta$-{P}u",
	author = "Olive, D. T. and Wang, D. L. and Booth, C. H. and Bauer, E. D. and Pugmire, A. L. and Freibert, F. J. and McCall, S. K. and Wall, M. A. and Allen, P. G.",
	journal = "J. Appl. Phys.",
	volume = "120",
	pages = "035103",
	year = "2016"
}

@ARTICLE{MM71,
	title = "The Length Changes in $\alpha$-Plutonium During Self-Irradiation Damage Below $4.2\,$K",
	author = "Marples, J. A. C. and Mortimer, M. J.",
	journal = "Phys. Lett.",
	volume = "34A",
	pages = "242--243",
	year = "1971"
}

@ARTICLE{K03,
	title = "Phase Transformations in Delta-Stabilized Plutonium",
	author = "Kitching, S. and Planterose, P. G. and Gill, D. C.",
	journal = "AIP Conf. Proc.",
	volume = "673",
	pages = "79--81",
	year = "2003"
}

@article{B07,
	title = "Self-irradiation effects in plutonium alloys",
	author = "Baclet, N. and Oudot, B. and Grynszpan, R. and Jolly, L. and Ravat, B. and Faure, P. and Berlu, L. and Jomard, G.",
	journal = "J. Alloys Compd.",
	volume = "444--445",
	pages = "305--309",
	year = "2007"
}

@INBOOK{H2,
	title = "Plutonium Handbook 2nd ed.",
	author = "Clark, D. L. and Geeson, D. A. and Hanrahan, R. J. Jr.",
	publisher = "Am. Nucl. Soc.",
	address = "Westmont, Ill.",
	volume = "2",
	pages = "1019",
	chapter = "12.3.4",
	note = "Fig. 24",
	year = "2019"
}

@INBOOK{H21,
	title = "Plutonium Handbook 2nd ed.",
	author = "Clark, D. L. and Geeson, D. A. and Hanrahan, R. J. Jr.",
	publisher = "Am. Nucl. Soc.",
	address = "Westmont, Ill.",
	volume = "2",
	pages = "1016",
	chapter = "12.3.3",
	note = "Fig. 21",
	year = "2019"
}

@INBOOK{H22,
	title = "Plutonium Handbook 2nd ed.",
	author = "Clark, D. L. and Geeson, D. A. and Hanrahan, R. J. Jr.",
	publisher = "Am. Nucl. Soc.",
	address = "Westmont, Ill.",
	volume = "2",
	pages = "1017",
	chapter = "12.3.4",
	note = "Fig. 22",
	year = "2019"
}

@INBOOK{H3,
	title = "Plutonium Handbook 2nd ed.",
	author = "Clark, D. L. and Geeson, D. A. and Hanrahan, R. J. Jr.",
	publisher = "Am. Nucl. Soc.",
	address = "Westmont, Ill.",
	volume = "2",
	pages = "1011--1012",
	chapter = "12.3.1",
	note = "Table 5",
	year = "2019"
}

@ARTICLE{C14a,
	title = "Nanoscale heterogeneity, premartensitic nucleation, and a new plutonium structure in metastable $\delta$ fcc Pu-Ga alloys",
	author = "Conradson, Steven D. and Bock, Nicolas and Castro, Julio M. and Conradson, Dylan R. and Cox, Lawrence E. and Dmowski, Wojciech and Dooley, David E. and Egami, Takeshi and Espinosa-Faller, Francisco J. and Freibert, Franz J. and Garcia-Adeva, Angel J. and Hess, Nancy J. and Holmström, Erik K. and Howell, Rafael C. and Katz, Barbara and Lashley, Jason C. and Martinez, Raymond J. and Moore, David P. and Morales, Luis A. and Olivas, J. David and Pereyra, Ramiro S and Ramos. Michael and Rudin, Sven P. and Villella, Phillip M.",
	journal = "Phys. Rev. B",
	publisher = "Am. Inst. Phys.",
	volume = "89",
	pages = "224102",
	year = "2014"
}

@ARTICLE{C14b,
	title = "Intrinsic nanoscience of $\delta$ Pu-Ga Alloys: Local Structure and Speciation, Collective Behavior, Nanoscale Heterogeneity, and Aging Mechanisms",
	author = "Conradson, Steven D. and Bock, Nicolas and Castro, Julio M. and Conradson, Dylan R. and Cox, Lawrence E. and Dmowski, Wojciech and Dooley, David E. and Egami, Takeshi and Espinosa-Faller, Francisco J. and Freibert, Franz J. and Garcia-Adeva, Angel J. and Hess, Nancy J. and Holmström, Erik K. and Howell, Rafael C. and Katz, Barbara and Lashley, Jason C. and Martinez, Raymond J. and Moore, David P. and Morales, Luis A. and Olivas, J. David and Pereyra, Ramiro S and Ramos. Michael and Jeffrey, H. Terry and Villella, Phillip M.",
	journal = "J. Phys. Chem. C",
	publisher = "Am. Chem. Soc.",
	volume = "118",
	pages = "8541--8563",
	year = "2014"
}

@ARTICLE{GS,
	title = "Effect of Intense Self-Irradiation on the Phase Composition of Metallic Plutonium",
	author = "Gorbunov, S. I. and Seleznev, A. G.",
	journal = "Radiochemistry",
	volume = "43",
	number = "2",
	pages = "111--117",
	year = "2001"
}

@ARTICLE{SSGS,
	title = "Ab initio calculations for void swelling in $\alpha$- and $\delta$-plutonium",
	author = "Sadigh, B. and S{\"o}derlind, P. and Goldman, N. and Surh, M. P.",
	journal = "Phys. Rev. Matl.",
	volume = "6",
	pages = "045005",
	year = "2022"
}

@ARTICLE{K93,
	title = "Microstructural Observations of $\alpha$-Quartz Amorphization",
	author = "Kingma, K. J. and Meade, C. and Hemley, R. J. and Mao, H.-K. and Veblen, D. R.",
	journal = "Science",
	volume = "259",
	pages = "666--669",
	year = "1993"
}

@ARTICLE{H94,
	title = "High-pressure behavior of silica",
	author = "Hemley, R. J. and Prewitt, C. T. and Kingma, K. J.",
	journal = "Reviews of Mineralogy",
	volume = "29",
	pages = "41--81",
	year = "1994"
}

@ARTICLE{MH72,
	title = "Length Changes in $\beta$-Plutonium due to Self-Irradiation Damage at 4K",
	author = "Marples, J. A. C. and Hall, R. O. A.",
	journal = "J. Nucl. Matl.",
	volume = "42",
	pages = "212--216",
	year = "1972",
	note = "Despite the title, this paper also discusses length changes of  the $\alpha$ and $\delta$ phases at 4 K"
}

@ARTICLE{H08,
	title = "Plutonium---An Element Never at Equilibrium",
	author = "Hecker, S. S.",
	journal = "Metal. and Matl. Trans. A",
	volume = "39A",
	pages = "1585--1593",
	year = "2008",
	note = "The reference for Table I should be to Marples and Hall J. Nucl. Matl. 42, 212."
}

@ARTICLE{CuZr,
	title = "Low temperature specific heat anomalies associated with the boson peak in CuZr-based bulk metallic glasses",
	author = "Li, Y. and Bai, H. Y. and Wang, W. H. and Samwer, K",
	journal = "Phys. Rev. B",
	volume = "74",
	pages = "052201",
	year = "2006"
}

@ARTICLE{CuZrTi,
	title = "Measurement of low temperature transport properties of Cu-based Cu-Zr-Ti bulk metallic glass",
	author = "Kuo, Y. K. and Sivakumar, K. M. and Su, C. A. and Ku, C. N. and Lin, S. T. and Kaiser, A. B. and Qiang, J. B. and Wang, Q. and Dong, C.",
	journal = "Phys. Rev. B.",
	volume = "74",
	pages = "014208",
	year = "2006"
}

@ARTICLE{Migliori,
	title = "Resonant ultrasound spectroscopic techniques for measurement of the elastic modulii of solids",
	author = "Migliori, A. and Sarrao, J. L. and Visscher, W. M. and Bell, T. M. and Lei, M. and Fisk, Z. and Leisure, R. G.",
	journal = "Physica",
	volume = "183B",
	pages = "1--24",
	year = "1993"
}

\end{document}